\documentclass[final]{vgtc}                          

\ifpdf
  \pdfoutput=1\relax                   
  \usepackage{graphicx}                
  \DeclareGraphicsExtensions{.pdf,.png,.jpg,.jpeg} 
\else
  \usepackage{graphicx}                
  \DeclareGraphicsExtensions{.eps}     
\fi%

\graphicspath{{figures/}{pictures/}{images/}{./}} 
\usepackage{microtype}                 
\PassOptionsToPackage{warn}{textcomp}  
\usepackage{textcomp}                  
\usepackage{mathptmx}                  
\usepackage{times}                     
\usepackage{cite}                      
\usepackage{tabu}                      
\usepackage{booktabs}                  

\onlineid{0}

\vgtccategory{Research}

\vgtcinsertpkg

\title{Anatomy Studio II \\
A Cross-Reality Application for Teaching Anatomy}

\author{\scriptsize Joaquim Jorge\thanks{e-mail: jorgej@ieee.org}\\ 
       \scriptsize Univ. de Lisboa \\ Portugal  %
\and \scriptsize Pedro Belchior\thanks{e-mail: pedro.belchior@tecnico.ulisboa.pt} \\
      \scriptsize Univ. de Lisboa \\ Portugal  %
\and \scriptsize Abel Gomes\thanks{e-mail: agomes@di.ubi.pt} \\
       \scriptsize Univ. da Beira Interior \\ Portugal 
       \and \scriptsize Maur\'icio~Sousa\thanks{e-mail: viva.o.mauricio@gmail.com}\\ %
     \scriptsize Univ. Toronto \\ Canada 
\and \scriptsize João Pereira\thanks{e-mail: jap@inesc-id.pt} \\%
       \scriptsize Univ. de Lisboa \\ Portugal %
       \and \scriptsize Jean-François Uhl\thanks{e-mail: jeanfrancois.uhl@gmail.com} \\ %
              \scriptsize Université de Paris \\ France} %

\teaser{
  \centering
  \includegraphics[width=\linewidth]{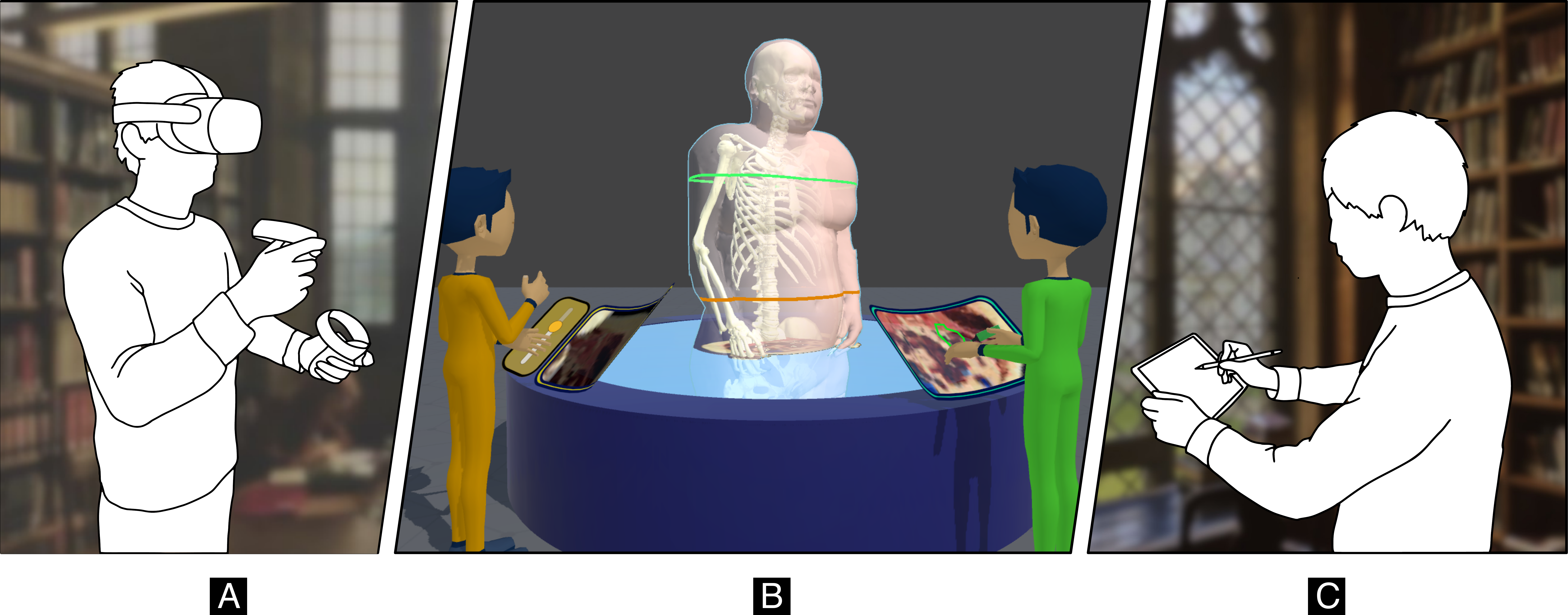}
  \caption{
  Anatomy Studio II illustrating the Cross-Reality use cases A) Immersed person wearing a VR Headset with controllers  B) Shared View of the Virtual Environment from a third person perspective  C) Connected user interacting with a tablet}
  \label{fig:teaser}
}

\abstract{
Virtual Reality has become an important educational tool, due to the pandemic and increasing globalization of education. 
In this paper, we present a framework for teaching Virtual Anatomy at the university level. 
Because of the isolation and quarantine requirements and the increased international collaboration, virtual classes have become a staple of today’s curricula. 
Our work builds on the Visible Human Projects for Virtual Dissection material and provides a medium for groups of students to do collaborative anatomical dissections in real-time using sketching and 3D visualizations and audio coupled with interactive 2D tablets for precise drawing. We describe the system architecture, compare requirements with those of a previous development 
\cite{ZORZAL201974} and discuss the preliminary results. Discussions with Anatomists show that this is an effective tool. We introduce avenues for further research and discuss collaboration challenges posed by this context.
} 

\CCScatlist{
  \CCScatTwelve{Human-centered computing}{Visu\-al\-iza\-tion}{Visu\-al\-iza\-tion techniques}{Treemaps};
  \CCScatTwelve{Human-centered computing}{Visu\-al\-iza\-tion}{Visualization design and evaluation methods}{}
}

\begin{document}

\maketitle


\section{Introduction}

The pandemic has drastically changed everyday life, changing the nature of traditional education from K12 to University levels. While Zoom and similar tools have emerged as a way to overcome the limitations posed by confinement and social distancing norms. However, two-way videoconferencing is a limited
tool for many medical and science disciplines that entail dealing with physical or 3D digital models especially in collaborative laboratory assignments. 

In this paper we explore Cross-Reality as a platform to support interaction between teachers, learners and 3D digital models, especially in the context of Digital Anatomy. Providing remote education through Augmented (AR) and Virtual Reality (VR) is a promising approach to this challenge. AR and VR enable instructors to conduct laboratory sessions remotely without travel and provide students with immersive experiences that mimic realistic scenarios. Current approaches, however, fail to deliver compelling experiences or offer beneficial interactions for either instructors or trainees alike because they focus on scripted game-like scenarios such as operating the instruments. There is no interaction with a remote instructor who could provide insights and correct errors. Other approaches focus on tele-operation and live assistance. No current system can deliver remote anatomy  training as a collaborative effort between professors and students. Our work aims to overcome this challenge. This paper focuses on Collaborative Extended Reality as a novel tool for Education in Digital Anatomy.

Digital Anatomy~\cite{DA_2021} has emerged as an essential subfield of Anatomy that processes the human body 
using computers. 3D reconstruction tools have been developed over the years, being always complementary to cadaver dissection.
Anatomy teachers' main goal is to provide a greater understanding of the spatial structures of the body and its internal organs thanks to 3D reconstruction techniques. 


The authors have previously worked on using Mixed reality for teaching performed a 3D model of the body from the slices of Korean Visible human (KVH) data~\cite{Park_2005}. 
The anatomical structures are precisely segmented manually using 2D images, then reconstructed and displayed as a 3D vector model. These volume models can then display an arbitrary section of the model and provide a virtual dissection function.  
Conventionally, this manual segmentation of anatomical slices process has been achieved with the Winsurf®\footnote{https://winsurf.software.informer.com/} software by producing an interactive 3D vector model of the whole body.
This is slightly different for the digital data from CT angiograms, where the segmentation process is fully automated by dedicated software. 
The process using Winsurf from anatomical slices has two main advantages:

\textbf{First}, to train the students in order to identify the anatomical structures on the 2D slices, and to help them visualize each structure in the 3D space. This 2D manual segmentation of each anatomical element could be considered as a computerized dissection. It can be followed and improved by the interactive display of the 3D object during the 2D segmentation process, enabling the representation of anatomical structures in the 3D space.

\textbf{Second}, it provides a 3D atlas of the whole body that can be used as a dissection learning tool (see Figure 1).
For this, the different anatomical elements are gathered into  3D vector models and exported to publishing software such as Adobe Acrobat®\footnote{https://www.adobe.com} 3D format, producing a 3D atlas of the whole body~\cite{Uhl_Mogorron_Chahim_2021}.

\textbf{Third}, this methodology of manual segmentation slice by slice to create the 3D objects is of high educational value because the outlining process can be followed step by step in real time with display of the resulting 3D object. Errors in segmentation and outlining can be corrected this way. In the same way, the confrontation of the different segmented objects makes it possible to correct the mistakes in the anatomical relation between different organs.
However, the process suffers from two major drawbacks. Indeed the tools used by this method work offline and do not support remote collaborative work. These shortcomings are a major motivator of our proposed approach.

\section{Related Work}
Currently, online learning is primarily limited to learning valuable skills (Udemy\footnote{https://www.udemy.com}, CreativeLive\footnote{https://www.creativelive.com}), completing multiple lectures (Coursera\footnote{https://coursera.org/}), or improving existing areas of knowledge (Pluralsight\footnote{https://www.pluralsight.com/}). 
Although there are more detailed learning experiences than elementary training (such as the Udacity\footnote{https://www.udacity.com} self-driving car course), most online learning has yet to provide a robust alternative to traditional on-site learning, and most of these sites focus on conventional interactive media, outside VR or XR. 
Indeed, previous mixed reality approaches, have contributed a variety of techniques for enhancing remote collaborative~\cite{gutwin2002descriptive, sousa2019warping}. 
In Negative Space~\cite{sousa2019negative}, the authors' explored how people and content should be presented for discussing 3D renderings within collaborative sessions. 
In a similar note, MAGIC~\cite{cat2020magic}, a novel approach to improving pointing agreement in shared 3D workspaces, communication through nonverbal cues while sharing the same perspective. 
Furthermore, there is a number of approaches to bring the benefits of novel interactive approaches such as AR and VR to applications in health and medical training. Existing approaches such as Shenai et al’s VIPAR~\cite{shenai2011virtual} system focused on assistance in medical procedures. 
Early work by Lee et al.~\cite{lan2007remote} used haptic feedback devices to deliver one-on-one training of operating surgical instruments.  McKnight et al.~\cite{mcknight2020virtual} showed that hard-mounted display hardware is constantly improving, software support for training and education is still lacking. Yet, many approaches using VR for surgical training provide game-like scripted scenarios. They do not leverage the rich interaction between instructors and trainees. 
Our vision is that collaborative XR environments can provide significant value added for University Curricula. 
For example, the entire Digital Anatomy curriculum could be created in a virtual environment through VR applications using Commercial Off-the-Shelf Affordable hardware such as Oculus Quest\footnote{https://www.oculus.com/quest-2/} or PICO\footnote{https://www.pico-interactive.com/}, allowing students to virtually attend lectures while maintaining a collaborative learning experience with other students. This immersive learning has many advantages, such as attending a virtual anatomy class where an instructor can provide assignments and students can virtually collaborate on these in a virtual laboratory.

\section{Our Approach}
\begin{figure}[!t]
  \resizebox{\columnwidth}{!} {
    \centering
    \includegraphics{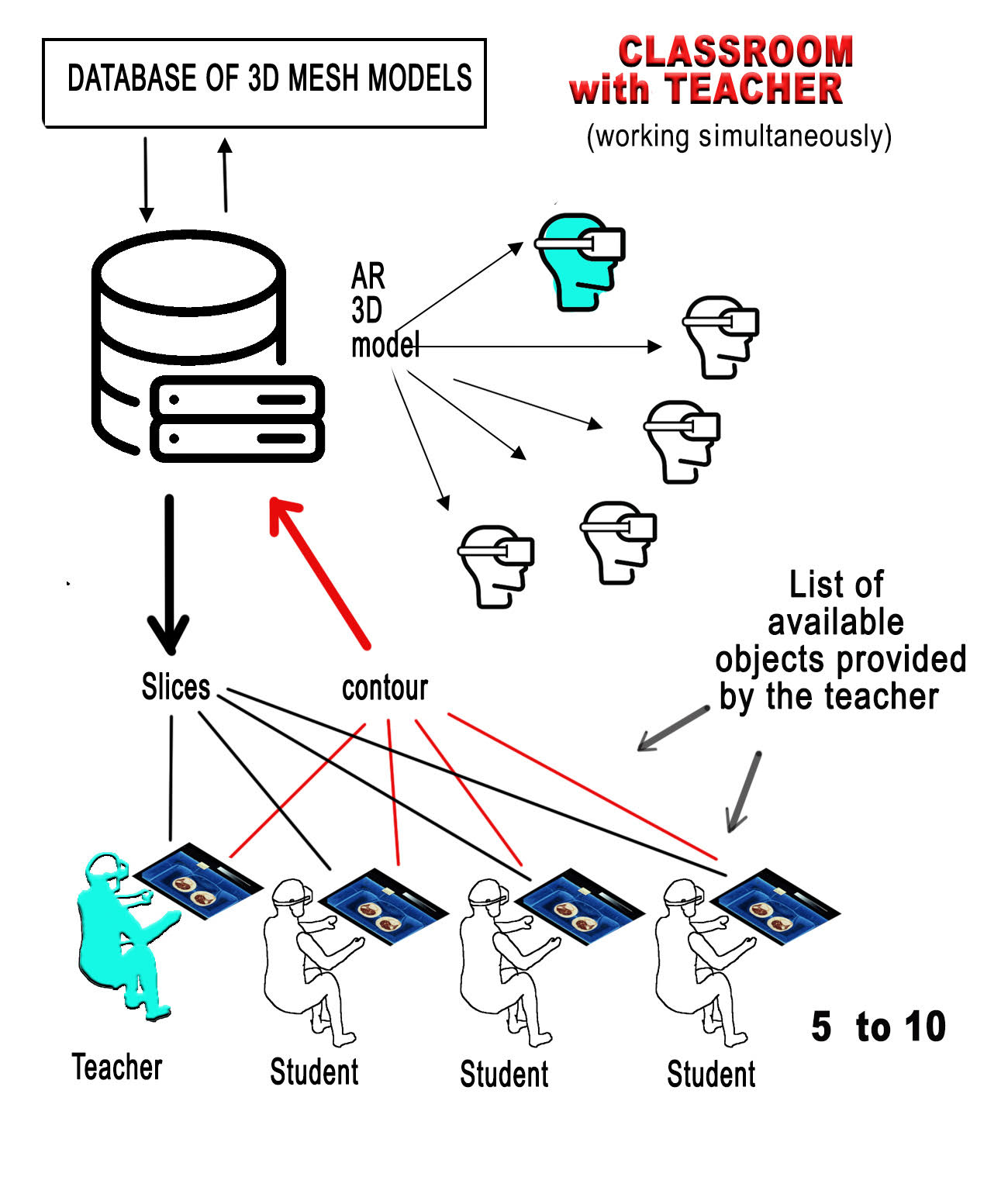}
    }
    \caption{Anatomy Studio II: using a Database of 3D Models and Digital Visible Human representations. Students can access information from the database and store digitized contours and 3d surface renderings of anatomical structures. The teacher can distribute lab assignments and oversee students working in remote locations.}
    \label{fig:sysarch}
\end{figure}

Contrary to Anatomy Studio I, that focused on same-place, same-time collaboration, we are interest on developing a distributed architecture for collaborative learning as depicted in Figure~\ref{fig:sysarch}. Anatomical slices from the visible female\footnote{https://www.nlm.nih.gov/databases/download/vhp.html} 
are used similarly to create 2D mesh models of the anatomical structures by manual outlining on tablets.



Different collaborative scenarios in real time can be achieved online: Group of students as illustrated in Figure~\ref{fig:teaser}, students with teacher (Figure~\ref{fig:sysarch}) or a students either working in groups or by themselves. In contrast, the conventional Anatomy classroom includes up to ten dissection tables, that accommodate 20-30 students as can be seen on this example~\footnote{https://www.youtube.com/watch?v=kqrZEzwqUEs}.
Note that our approach is focused on having students working in separate virtual lab rooms. This is supported by a client-server architecture as described in what follows.



The client applications are the user-facing portion of Anatomy Studio. They provide the user-interface for browsing the anatomical datasets, contouring anatomical structures, visualizing the reconstructed anatomical volumes and collaborating with other users. 
Anatomy Studio II provides Client Applications for multiple devices, including desktop computers, smartphones, tablets and standalone VR headsets, implementing different user input methods – 
including speech, touch gestures and remote controller and hand tracking when interacting with Head-Mounted Displays (HMD). 


The server application is responsible for synchronizing the state of each user’s avatar, dataset navigation, ongoing contouring work and other data across all client applications that are authorized to view that data – i.e. for all other users that are on the same virtual laboratory session.
The server application is also responsible for all communications with the database.






The Dataset Repository is run as 
an HTTP file service, providing the client applications with access to the slice datasets used by Anatomy Studio. The slice datasets take up between 100 and 200 GB each, and would be impractical to ship with the client applications. To reduce the bandwidth usage on the server, and provide a smoother user experience, the slice datasets are tiled at various zoom levels. 



\subsection{Cross-Reality Collaboration}
We propose to explore remote training in virtual dissection. Our approach blends the instructor and trainee environments through AR and VR using synthetic avatars and sketch-based anatomical reconstruction. Educators and students will be in separate environments, equipped with tablet devices or VR HMDs. Our system enables immersed participants to see each other as if they were in the same physical space. Thus our research facilitates immersive learning experiences through novel interactive tools for annotation and communication between instructors and trainees. Furthermore, since instructors need not be present constantly, primarily for questions and monitoring, Anatomy Studio II (ASII) can serve students in multiple remote locations.
ASII supports cross-reality interaction so that people operating different devices share the same application state. All users have an avatar in ASII virtual environment. Those running a VR client application can directly control the movement of their avatars. Additionally, users working at a desktop or mobile client application can move their avatars according to their actions using pre-programmed animations.
Users can talk to, and listen to, other users in the same session.
ASII distinguishes different participants using color-coded avatars.
Each participant in a lab session can see which slices others are looking at by looking for the slices marked with that user's color on the dataset to facilitate coordination. Furthermore, ASII provides navigation elements such as matching color-coded lines that appear on the cadaver's 3D model to depict the slice each user is looking at. Finally, people can filter user-created contours and volumes to identify participants or anatomical structures. Users can have existing contours and volumes visible while they create new contours to serve as guides and have any collisions between the two visually highlighted. 
Participants can grade 
others' work using a 5-star grading system and check the accuracy of their work against an expert-created atlas to support gamification.

\subsection{Sketching Contours and Reconstructions}


As we have seen above, we support the teaching of anatomy through virtual dissection. To this end, we use a kind of one-directional scanning of Visible Human (VH) slices to allow the students to edit and reconstruct anatomical structures interactively slice-by-slice, forward and backward. This slice-by-slice procedure is inspired in the Boissonnat work \cite{1986-Boissonnat}. Furthermore, this interactive forward-backward procedure enables students to keep the connectivity 
of the reconstructed anatomical surfaces through two steps: (i) to find the contours of each anatomical structure in a slice and its next slice; (ii) to extend a polygonal mesh between adjacent contours. 

\begin{figure}[t!]
  \resizebox{\columnwidth}{!} {
    \centering
    \includegraphics{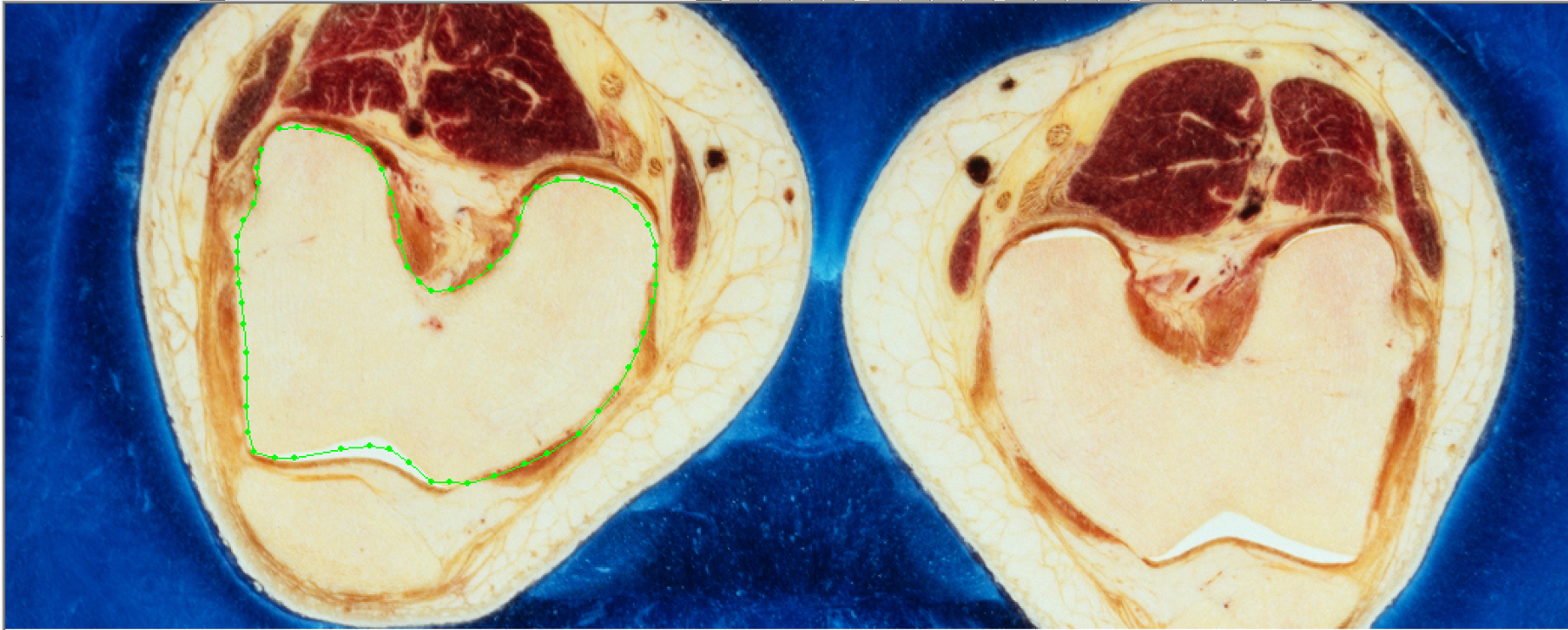}
    }
    \caption{Example Visible Human Slice with Contours highlighted}
    \label{fig:slices}
\end{figure}

\begin{figure}[b!]
  \resizebox{\columnwidth}{!} {
    \centering
    \includegraphics{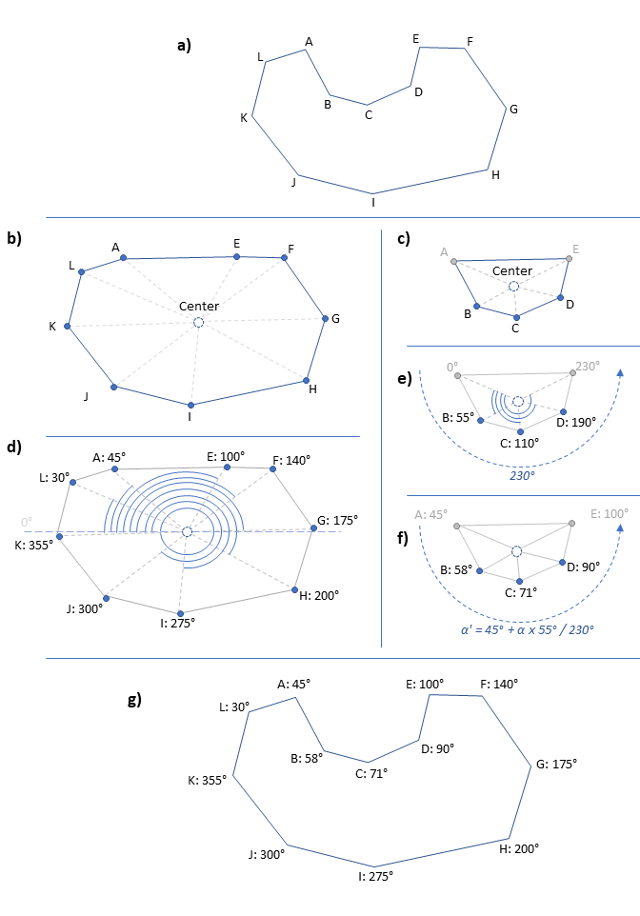}
    }
    \caption{Contour Processing from Sketched Data}
    \label{fig:contours}
\end{figure}

Reconstructing the 3D volume of an anatomical structure consists of drawing planar contours over a set of parallel anatomical slices and then rebuilding the 3D mesh bound by those contours. We do so by first ordering the contours along the direction defined by the normal to the planes containing the contours, then building all the 3D shells defined by each pair of consecutive contours, and finally merging all 3D shells to form the complete volume.

Building the 3D shells is, however, not trivial. A naive approach would be to, given a pair of non-coplanar contours, pick a point on one of them, find the closest point on the other, and create a triangle strip from there by alternatively moving along each contour, in the same clockwise direction, one vertex at a time. However, since the contours can have different shapes, numbers of vertices, and varying degrees of overlapping, such naive approaches will result in inaccurate 3D shells. So, building the shell first requires that contours be normalized somehow.

To do this, we created the algorithm described in Figure~\ref{fig:contours}. First, we calculate the convex hull of the contour. Then, we find the hull's center and calculate the angles between the X-axis and each of the lines connecting the hull's center to each hull vertex. We then repeat the process for each hole and sub-hole of the contour, should they exist, with alternating winding orders. The result is a contour whose vertices span,  clockwise, increasing angles from 0 to 360 degrees. 

\begin{figure}[t!]
    \centering
    \includegraphics[width=\linewidth]{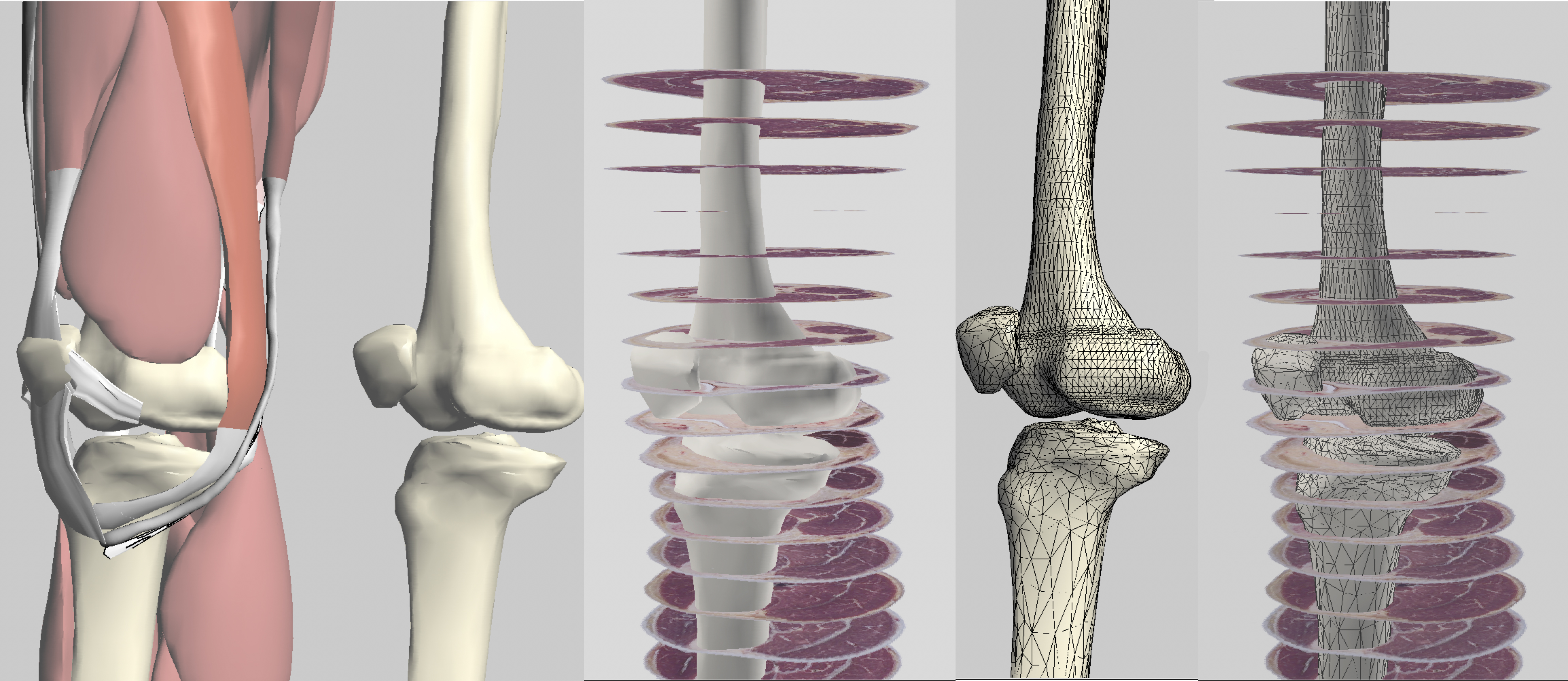}
    \caption{3D vector mesh models of the right knee.}
    \label{fig:montage}
\end{figure}

Once we have processed two non-coplanar contours in this manner, we can create the 3D mesh joining them. To this end, we traverse the lists of vertices of both contours, ascending order of angle-values, starting with the lowest angle-value vertices of each contour and then picking the next lowest angle-value vertex from either contour, forming a triangle with it. The process is repeated until we have visited all vertices. 3D structures are thus created by connecting adjacent contours in a process similar to that applied in Anatomy Studio I~\cite{ZORZAL201974}. The process is illustrated in Figure \ref{fig:montage}.

\section{Discussion}


Anatomy studio's performance can be hindered by two main bottlenecks - the client application's graphical output and network bandwidth. Lower spec devices, such as older tablets or smartphones, may have trouble rendering more complex 3D models, resulting in stuttering, low frame rates, and input lag. In addition, anatomy Studio sessions with too many users could cause the network traffic to quickly grow exponentially, overwhelming the bandwidth of either client devices or the Anatomy Studio server, or both.  To tackle these two bottlenecks, we decided on the primary strategy, dividing users into groups.
    Users first join a session. Therefore, users in one session cannot group with, interact with, or see users' work in other sessions. 
While in a session, we join users in groups of three to four. This can be voluntary or automatic, depending on 
session setup. 

Dividing users into groups like this serves both human and system performance purposes. By limiting users to interacting only with those other users in the same session and group, we lower the cognitive load and the potential for distraction for all of them and prevent any crowding issues around the virtual table in VR space. Nevertheless, we also reduce the amount of data rendered on the client devices and the number of user actions that have to be propagated to other users through the network. Instead of growing exponentially, network traffic now grows linearly with users. We have implemented other performance strategies for more specific cases, such as pre-rendering full-body views of the cadaver for low-performance devices.

\section{Conclusions}

We have presented Anatomy Studio II, a cross-reality application to support the teaching of Digital Anatomy.
This article considers Cross Reality as a platform to support interactions between teachers, learners, and 3D digital models, especially in the context of digital anatomy. Providing distance learning via augmented reality (AR) and virtual reality (VR) is a promising approach to this challenge. With AR and VR, instructors can conduct lab sessions remotely without traveling and provide students with an immersive experience that mimics realistic scenarios. However, the current approach focuses on scripted game-like scenarios such as playing musical instruments, providing a compelling experience and valuable interactions for trainers and trainees. Unfortunately, you cannot do it. There is no interaction with remote teachers to provide insights and correct mistakes. Other approaches focus on teleoperation and live assistance.
Furthermore, current systems cannot provide remote anatomy training as a collaborative effort between professors and students. Our job is to address this challenge. This paper focuses on collaborative augmented reality as a new tool for digital anatomy education. Remote Collaborative Education via XR provides an approach that allows better use of precious human capital, improves the efficient use of scarce qualified resources, and makes other people more qualified via training and real-time remote teaching.

\acknowledgments{
This work was supported in part by the Portuguese Govt. \textit{Fundação para a Ciência e a Tecnologia}, under project UIDB/50021/2020.
}

\bibliographystyle{ieeetr}

\bibliography{main}
\end{document}